# Deformed square resonator lasers for optical frequency comb generation

Hai-Zhong Weng, Yong-Zhen Huang,* Yue-De Yang, Xiu-Wen Ma, Jin-Long Xiao, and Yun Du

*State Key Laboratory on Integrated Optoelectronics, Institute of Semiconductors,*

*Chinese Academy of Sciences, Beijing, 100083, China*

Deformed square resonators with the flat sides replaced by circular sides are proposed and demonstrated to enhance mode $Q$ factors and adjust transverse mode intervals using the regular ray dynamic analysis and numerical simulations. Dual-transverse-mode emissions due to the ultrahigh-$Q$ factors with different wavelength intervals are realized experimentally for AlGaInAs/InP circular-side square microlasers, and the stationary condition of the dual-mode emission is satisfied because the high-$Q$ confined modes have totally different mode numbers. Furthermore, optical frequency combs are generated using the dual-mode lasing microlaser as a seeding light source by cascaded four-wave mixing in a highly nonlinear optical fiber.

PACS: 42.55.Sa, 42.55.Px, 42.60.Da, 42.25.-p,

High quality ($Q$) whispering-gallery mode (WGM) microcavities exploiting total internal reflection (TIR) of light ray have attracted great attention for fundamental physics studies and applications of optoelectronic devices [1-8]. Low threshold directional emission microlasers have been realized using deformed microdisk resonators [9-14], and parametric frequency conversions in high-$Q$ microresonators have been applied to generate optical frequency comb (OFC) [15-18]. OFCs with equidistant spectral lines can provide unprecedented precision for optical frequency measurements and enable various applications in atomic clocks [19, 20], gas sensing, molecular fingerprinting [21, 22], optical waveform generation [23], and advanced modulation communications [24], etc. OFC was traditionally generated using femtosecond mode-locked lasers [19, 25, and 26]. The generations of OFC in ultrahigh-$Q$ microresonator have triggered intensive researches of four-wave mixing (FWM) in the microresonator under a continuous-wave laser pumping [27-29], and cascaded FWM in the nonlinear optical fiber has also been utilized for generating frequency combs via the pumping of two single-frequency lasers [30-32].

Recently, square resonators have received a great deal of interest for single-mode lasing and enhancement of mode $Q$ factors [33-36]. Dual-mode lasing with a tunable wavelength interval was demonstrated for square resonator microlasers with a patterned electrode and a vertex waveguide [33]. The large square resonators with a vertex waveguide can have high passive mode $Q$ factors and result in dual-mode lasing simultaneously for the fundamental and first-order transverse modes, as they have near equal mode $Q$ factors accounting the internal absorption loss. However, the mode $Q$ factors are greatly limited by the inevitable radiation loss from the vertices of the square resonator [33, 34]. In this letter, we introduce a novel circular-side square resonator (CSR) to modify the mode field patterns and enhance mode $Q$ factors. The concave mirrors are widely used in conventional gas or solid laser systems for focusing the light beam [37-39], but are rarely applied to the polygonal microresonators. Here, dual-mode lasing due to the high-$Q$ factors with changeable mode intervals is demonstrated by varying the radius of the circular-side. Frequency combs are obtained in nonlinear fiber using the dual-mode lasing microlaser as a seeding light source, which opens a way to realize OFC with a large line interval by simply adjusting the resonator form instead of the size.

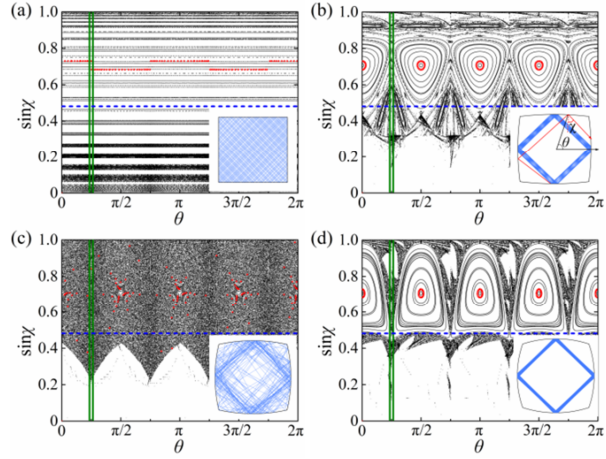

FIG. 1 The upper Poincaré surface of sections for the CSRs with the side length $a = 16$ μm at the deformation amplitude $\delta$ of (a) 0, (b) 0.5μm, (c) 0.9 μm, and (d) 1.3 μm, where $\theta$ and $\chi$ denote the polar angle of reflection position and incident angle as shown in the inset of (b). The horizontal dashed lines indicate the critical angle of the total internal reflection. Insets: The real ray trajectories with 200 bounces inside the CSR with the starting point at $\sin\chi = 0.73$ and $\theta = 0$, corresponding to the red symbols in the phase space.

The two-dimensional (2D) CSR is firstly simulated by the light ray analysis, with the refractive indices of 3.2 and 1.54 for the CSR and surrounded bisbenzo cyclobutene (BCB), respectively. Upper half Poincaré surface of sections with 360 initial polar angles and 2000 bounces are presented in Figs. 1(a)-1(d) for the CSRs with deformation amplitude $\delta$ of 0, 0.5, 0.9, and 1.3 μm, respectively, which is defined by $\delta = r - \sqrt{r^2 - (a/2)^2}$, where $a$ is the flat-side length and $r$ is the radius of the circular side. Light rays are approximately considered to be leakage as they impinge the vertices. The real ray trajectories trapped by TIR are illustrated in the insets with the starting point at $(\theta, \sin\chi) = (0, 0.73)$ and marked by the red symbols in the phase space. In Fig. 1(a), the horizontal lines in the phase space indicate that the mode light rays impinge whole resonator with the complementary incident angles at the adjacent sides. In Fig. 1(b), the intensive closed curves around $\theta = 0$, $\pi/2$, $\pi$, $3\pi/2$, and $2\pi$ are corresponding to high-$Q$ WGMs, surrounded by the external chaotic regions. At $\delta = 0.9$ μm, the ray orbits are scattered randomly with very small islands in Fig. 1(c), but regular triangle-like islands are formed again in Fig. 1(d) at $\delta = 1.3$ μm. The results indicate that the circular arcs can confine the mode light rays away from the vertices of the CSR at $\delta = 0.5$ and 1.3 μm, which can enhance the mode $Q$ factors and increase the transversal mode intervals due to the narrow light beams as shown in the insets.

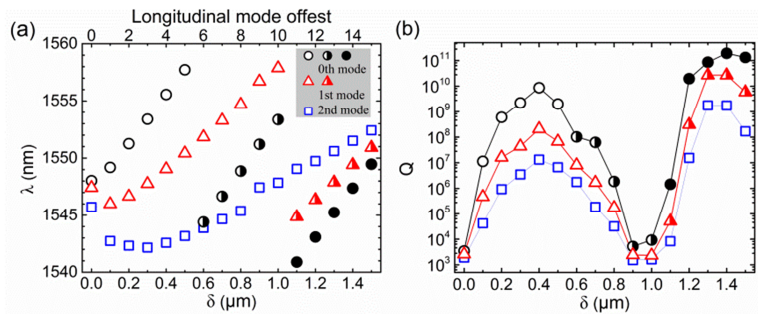

FIG. 2 (a) Resonance wavelengths around 1550 nm and (b) corresponding $Q$ factors versus the deformation amplitude $\delta$. The variation of the longitudinal mode number is also traced in (a) for the square marked modes, and the open, half-filled, and solid symbols belong to different longitudinal modes at a certain $\delta$. The fundamental (0th-order), first-order (1st-order), and second-order (2nd-order) transverse modes are marked by circle, triangle, and square symbols, respectively.

Then the mode characteristics are investigated by solving the wave equation using finite element method (FEM, COMSOL Multiphysics 5.0), for the 2D CSRs with an output waveguide at one vertex, marked by the vertical solid lines in Fig. 1. Taking $a = 16$ μm and the output waveguide width $w = 1.5$ μm, we simulate the transverse electric (TE) modes for the CSRs in the x-y planer with the z-direction magnetic field $H_z$ satisfying symmetric condition relative to the output waveguide. Resonance wavelengths and Q factors of the symmetric TE modes around 1550 nm versus deformation amplitude $\delta$ are plotted in Figs. 2(a) and 2(b), where the variation of the longitudinal mode numbers is shown in the upper x-axis for the modes marked by square symbols. The mode wavelengths are red-shift about 17 nm as $\delta$ increases 0.1 μm for the 0th-order transverse mode and the transverse mode intervals gradually increase with the increase of $\delta$ due to the effective resonance length differences for different transverse modes. The wavelength interval between the 0th-order and the 2nd-order transverse modes is almost equal to the longitudinal mode interval at $\delta = 0.6$ μm, as shown in Fig. 2(a) with the intersection between the half-filled circle and the open square symbol. As depicted in Fig. 2(b), the Q factors of the 0th-order, 1st-order, and 2nd-order transverse modes are $3.5\times10^3$, $2.6\times10^3$, and $2\times10^3$, respectively, as $\delta = 0$, then rapidly increase to $8.7\times10^9$, $2.1\times10^8$, and $1.3\times10^7$ as $\delta = 0.4$ μm, and drop to $5.2\times10^3$, $2.4\times10^3$, and $1.5\times10^3$ as $\delta = 0.9$ μm, but increase to the maximum values of $1.9\times10^{11}$, $2.7\times10^{10}$, and $1.8\times10^9$ again as $\delta = 1.4$ μm. For a circular resonator with a radius of 8 μm calculated in the same conditions, the mode Q factors are $1.9\times10^{12}$ and $6.1\times10^{11}$ for WGMs $TE_{190,\,1}$ and $TE_{178,\,2}$ at 1556.7 and 1549.6 nm, which are limited by the numerical computation precision. However, by connecting a 1.5 μm-wide waveguide to the circular resonator, the highest Q factor around wavelength 1550 nm is only $1.5\times10^4$ for a coupled mode at 1564.8 nm. For the CSR connecting four 1.5 μm-wide output waveguides at the vertices, the fundamental transverse mode at 1547.4 nm can still have high-Q factor of $1.5\times10^{11}$ as $\delta = 1.4$ μm and $a = 16$ μm. The ultrahigh-Q factors can be obtained for the CSR connected with four waveguides as the mode field distributions away from the vertices. So ultrahigh-Q confined modes can be maintained for two-dimensional integrated CSRs.

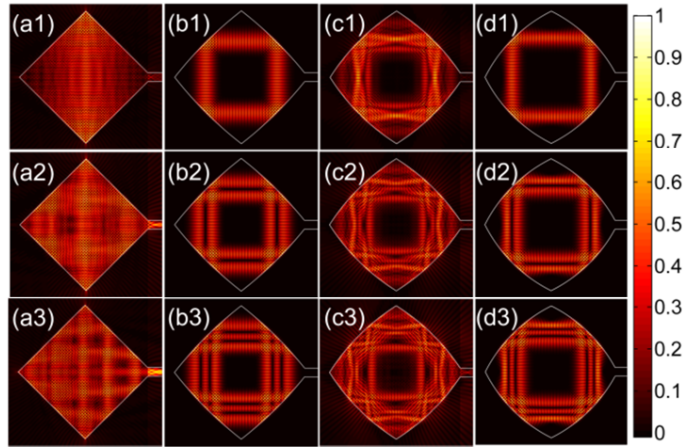

FIG. 3 Calculated magnetic field component $|H_z|$ for the 0th-order, 1st-order, and 2nd-order symmetric transverse modes are presented in (a1)-(a3) at $\delta = 0$, (b1)-(b3) $\delta = 0.5$ μm, (c1)-(c3) $\delta = 0.9$ μm, and (d1)-(d3) $\delta = 1.3$ μm, respectively.

The magnetic field distributions of $|H_z|$ for the 0th-order, 1st-order, and 2nd-order transverse modes are presented in Fig. 3 as $\delta = 0$, 0.5, 0.9, and 1.3 μm, respectively, with those on the right side including the output waveguide magnified by two times. For the flat-side square resonator, the modes have strong coupling efficiency to the output waveguide and large radiation loss in the vertices of the square as shown in Figs. 3(a1)-3(a3). At $\delta =$

0.5 μm, the mode fields are well confined in the transverse direction as shown in Figs. 3(b1)-3(b3), with near zero radiation loss at the vertices. At $\delta$ = 0.9 μm, the mode fields in Figs. 3(c1)-3(c3) distribute over the whole resonator again and have a large radiation loss, corresponding to the minimum mode $Q$ factors in Fig. 2(b). At $\delta$ = 1.3 μm, the mode fields are compressed again with a more narrow transmission path as shown in Figs. 3(d1)-3(d3).

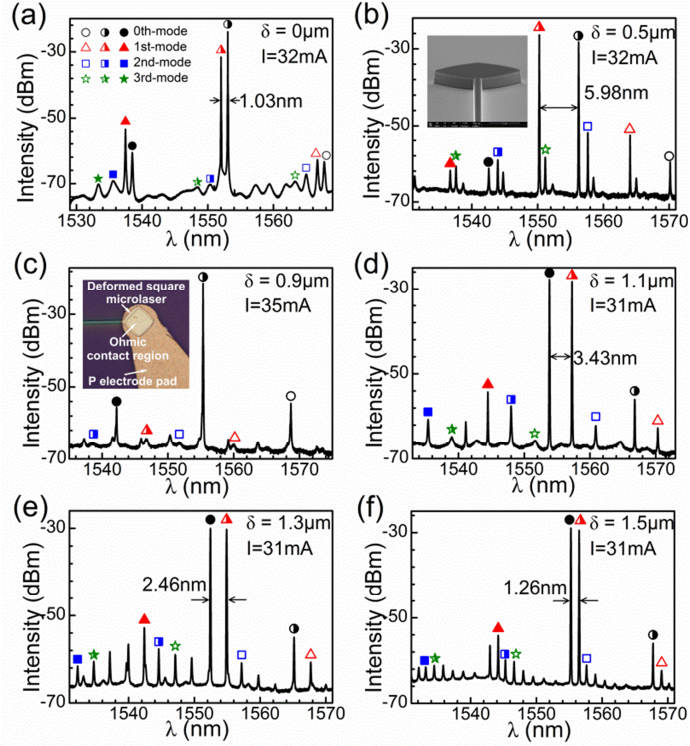

FIG. 4 (a)-(f) Lasing spectra for the CSR lasers with $\delta$ of 0, 0.5, 0.9, 1.1, 1.3, and 1.5 μm, respectively. The 0th-order, 1st-order, and 2nd-order, and third-order (3rd-order) transverse modes are marked by circle, triangle, square, and pentagram symbols, respectively, with the open, half-filled, and solid symbols belonged to different longitudinal modes. Insets in (b) and (c) are a side-view SEM image after ICP etching and a top-view microscope image of a microlaser.

The CSR microlasers with $a$ = 16 μm, $w$ = 1.5 μm are fabricated using an AlGaInAs/InP wafer with eight pairs of compressively strained multiple-quantum-wells as in [34]. The side-view scanning electron microscope (SEM) image after inductively coupled-plasma (ICP) etching is shown in the inset of Fig. 4(b), and top-view microscope image of a microlaser in the inset of Fig. 4(c). The CSR lasers are tested at the temperature of 288 K by butt-coupling a fiber to the cleaving facet of the output waveguide. Lasing spectra for the CSR microlasers are plotted in Figs. 4(a)-4(f) at $\delta$ = 0, 0.5, 0.9, 1.1, 1.3, and 1.5 μm, respectively. With the help of the numerical results, the 0th-order, 1st-order, and 2nd-order, and 3rd-order transverse modes are assigned and marked by the circle, triangle, square, and pentagram symbols, respectively. Dual-mode lasing with the wavelength intervals of 5.98, 3.43, 2.46 and 1.26 are obtained at $\delta$ = 0.5, 1.1, 1.3 and 1.5 μm, and near dual-mode lasing with an intensity ratio of 7.9 dB is observed at $\delta$ = 0. But single-mode lasing at 1555 nm with the side-mode-suppression-ratio (SMSR) of 33 dB is obtained for the CSR laser with $\delta$ = 0.9 μm at 35 mA. The results indicate that the CSRs with ultrahigh-$Q$ modes are easy to realize dual-mode lasing and the transverse mode interval can be adjusted by the radius of the circular sides. The transverse mode intervals $\Delta\lambda_{01}$, $\Delta\lambda_{02}$ and $\Delta\lambda_{03}$ are varied with $\delta$ at the rates of 5.2, 10.2 and 14.9 nm/μm for the CSR lasers as 0.5 ≤ $\delta$ ≤ 1.5 μm, which are in good agreement with the numerical results of 5.2, 9.8, and 14.1 nm/μm.

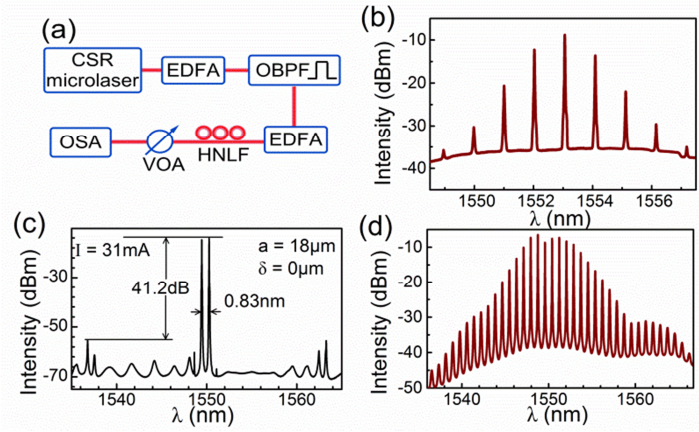

FIG. 5 (a) Schematic of the experiment setup used for generating OFC, where the CSR microlaser servers as seeding light source and lunches into the HNLF. EDFA = Erbium-doped fiber amplifier, OBPF = optical band-pass filter, HNLF = highly nonlinear fiber, VOA = variable optical attenuator, OSA = optical spectrum analyzer. (b) The OFC generation based on the microlaser in Fig. 4(a). (c) and (d) are the lasing spectrum and generated OFC spectrum for a microlaser with $a$ = 18 μm and $\delta$ = 0 at 31 mA.

Finally, the dual-mode-lasing microlasers are applied to produce OFC with the schematic diagram shown in Fig. 5(a). The dual-wavelength output is firstly pre-amplified by an EDFA and filtered by an OBPF. Then the filtered output is boosted up to ~500 mW with another EDFA and fed into a 1-km-long HNLF to initiate the cascaded FWM processes. The generated OFC is measured by an OSA after a VOA. The HNLF has a nonlinear coefficient of 10 $(W·km)^{-1}$ and a zero dispersion wavelength at 1535 nm, with the dispersion and dispersion slope of 0.389 ps/(nm·km) and 0.017 ps/($nm^2$·km) at 1550 nm. The OFC obtained based on the lasing spectrum in Fig. 4(a) is depicted in Fig. 5(b), and the lasing spectrum and generated OFC spectrum are shown in Figs. 5(c) and 5(d) for a square laser with $a$ = 18 μm and $\delta$ = 0 at 31 mA. The OFCs with nine and thirty seven spectral lines at the mode spacing of 1.03 and 0.83 nm are attained for the microlasers with $a$ = 16 and 18 μm. However, the CSR with ultrahigh-$Q$ factors will reduce the output power greatly, which limits its application for the OFC generation. So mode $Q$ factor and output efficiency should be compromised to reach dual-mode lasing and high output power. A pair of FWM peaks is clearly exhibited in the lasing spectra in Figs. 4(a) and 5(c), which indicates the coexistence of the dual-mode lasing. To avoid the inevitable mode competition in the bulk or quantum well semiconductor lasers, quantum dot semiconductor lasers were proposed to realize a stable dual-mode emission microlaser [40]. In the square resonator, the mode numbers $p$ and $q$ are the numbers of wave nodes along the directions of the square sides [41], and the high-$Q$ confined modes usually have different mode numbers $p$ and $q$, even they have the same longitudinal mode number or transverse mode number. So the ratio of the self-saturation coefficient to the cross-saturation coefficient will reduce from 4/3 for the same transverse mode to 8/9 for different transverse modes [42] in the square resonator microlasers, which satisfies the condition of the dual-mode stationary solution [40].

In summary, we have introduced a novel circular-side square resonator and simulated mode characteristics by the light ray analysis and the finite element methods. Benefitting from the enhanced mode $Q$ factors and controllable mode intervals, the resonators enable the dual-transverse-mode lasing with specific mode intervals, which are confirmed from the fabricated AlGaInAs/InP microlasers. Furthermore, the generations of optical frequency combs are demonstrated using the dual-mode lasing microlaser as a seeding light source in nonlinear fiber. The ultrahigh-$Q$ factors provide plenty space to adjust the mode $Q$ factor for improving the output efficiency.

The modes with the field patterns deviated from the vertices can still have very high-$Q$ factors as connecting four output waveguides to the four vertices of the resonator. We expect that the circular-side square microresonator will be a versatile microresonator in photonic integrated circuits for optical signal processing, optical delay and storage.

We thank the anonymous referee's comment on the stability condition for the dual-mode emission, and Prof. Qinghai Song for discussions and suggestions. This work is supported by the National Natural Science Foundation of China under Grant Nos. 61527823, 61235004, and 61321063.

------------------------------------

*yzhuang@semi.ac.cn